\documentclass[conference]{IEEEtran}
\IEEEoverridecommandlockouts
\usepackage{algorithmic}
\usepackage{graphicx}
\usepackage{balance}
\usepackage{epstopdf}
\usepackage{hyperref}
\usepackage{amsmath,bm}
\usepackage{amsfonts}
\usepackage{amssymb}
\usepackage{multirow}
\usepackage{subfigure}
\usepackage[inline]{enumitem}
\usepackage{xcolor}
\usepackage{float}
\usepackage[flushleft]{threeparttable}
\usepackage{algorithm}
\usepackage{nomencl}
\usepackage{authblk}
\usepackage{comment}
\usepackage{array}
\PassOptionsToPackage{hyphens}{url}
\usepackage{hyperref}

\newcommand{\eat}[1]{}

\begin{document}

\title{Detecting Hidden Attackers in Photovoltaic Systems Using Machine Learning}

\author[1,2]{Suman Sourav}
\author[2]{Partha P. Biswas}
\author[1,2]{Binbin Chen}
\author[2]{Daisuke Mashima}

\affil[1]{Singapore University of Technology and Design, Singapore; e-mail: [suman\_sourav, binbin\_chen]@sutd.edu.sg}
\affil[2]{Advanced Digital Sciences Center, Singapore; e-mail: [partha.b, daisuke.m]@adsc-create.edu.sg} 

\IEEEoverridecommandlockouts

\maketitle

\IEEEpubidadjcol

\begin{abstract}
In modern smart grids, the proliferation of communication enabled distributed energy resource (DER) systems has increased the surface of possible cyber-physical attacks. Attacks originating from the distributed edge devices of DER system, such as photovoltaic (PV) system, is often difficult to detect. An attacker may change the control configurations or various setpoints of the PV inverters to destabilize the power grid, damage devices, or for the purpose of economic gain. A more powerful attacker may even manipulate the PV system metering data transmitted for remote monitoring, so that (s)he can remain hidden. In this paper, we consider a case where PV systems operating in different control modes can be simultaneously attacked  and the attacker has the ability to manipulate individual PV bus measurements to avoid detection.
We show that even in such a scenario, with just the aggregated measurements (that the attacker cannot manipulate), machine learning (ML) techniques are able to detect the attack in a fast and accurate manner. %
We use a standard radial distribution network, together with real smart home electricity consumption data and solar power data in our experimental setup. We test the performance of several ML algorithms to detect attacks on the PV system. Our detailed evaluations %
show that the proposed intrusion detection system (IDS) is highly effective and efficient in detecting attacks on PV inverter control modes.

\end{abstract}

\section{Introduction}

The fast depletion of fossil fuels, environmental regulation and imposition of carbon tax in some cases compelled many countries to quickly adopt alternate and replenishable sources of energy. Solar energy, a form of renewable energy, is harnessed by the well-established PV system technology. Globally, the growth in PV system installed capacity has been phenomenal in the last couple of decades, %
and just in the year 2019, about 109GW PV capacity was installed~\cite{IEA}. Together with the high penetration of renewable energy, modern grids are accommodating internet-of-things (IoT) devices for efficient control and management of the grid. Though sophisticated information technology network facilitates faster communication and overall performance enhancement of the grid, it also opens up channels for cyberattackers to intrude into the power system and perform any offensive manoeuvre that makes the system behave undesirably.

Grid or transmission system operator usually do not have visibility of DER plants (e.g., solar plants), and in most cases the monitoring of the plant is restricted to the local distribution substation level. Furthermore, renewable sources are inherently variable. Limited visibility as well as uncertainty in renewable sources can easily be exploited by an attacker to launch attacks through the edge devices of the PV systems~\cite{shilay2017catching}. The variability in system performance and output would even make it harder to distinguish between an actual attack and a regular change in system parameter. Attack on PV system control characteristics may jeopardise the system voltage. The impact of the attack would be severe if a majority of the PVs connected to the system are manipulated (to give reduced output or fully disconnected) or the attack is launched during peak-loading scenarios of the network. Such an attack would have economic consequences as well. Lowering the PV system active power setpoint would force the consumers to draw more active power from the thermal generator(s) connected to the grid, thereby increasing both fuel cost and emission. As an example, a change of setpoint even by 0.1 MW (lower) would incur an additional fuel cost of anywhere between \$0.175 to \$0.325 per hour depending on the type of fossil fuel used by the generator \cite{BISWAS20171194}. The cost follows quadratic relationship with the active power, implying even more commercial impact if the setpoint is lowered further.

In a distribution network, different PVs can operate in different control modes, and they can be attacked simultaneously. Most of the existing studies \cite{shilay2017catching,li2020data, qi2016cybersecurity} considered attack only on single mode of operation, i.e., where all the PV systems are operating in the same mode. In contrast, here we consider and aim to highlight the case of a combined attack on multiple, different modes. We discuss the impact of such an attack and propose a method to detect it. Moreover, we consider a sophisticated attacker who tries to stay hidden by manipulating the PV bus measurements, i.e., by sending PV bus measurements as if there is no attack. {Such hidden attackers often target to gain information by observing the system and/or to disrupt the system on a specific planned time.}

A cyber rule-based intrusion detection system (IDS) might not be able to detect such attacks on the edge devices of PV system. Therefore, we need to leverage physical system data to analyze and detect. The adoption of machine intelligence in detecting cyberattacks in smart grid on various aspects is widespread. Several machine learning (ML) algorithms have also been tested in the context of PV systems~\cite{shilay2017catching,li2020data}. In a supervised ML method, the algorithm is trained with system-wide data for both normal and attack scenarios in the network considering variable load demand and solar power. Based on the training imparted, the algorithm is expected to classify a new set of data in the category of normal or attacked data. 
The method is fast and quite accurate. Also, the ML-based detection algorithm works in a non-invasive manner as it does not interfere with the normal power system operation. In a power grid, measurement data from some metering instruments might be missing occasionally due to communication loss or faulty instruments. A trained ML-model might still be able to correctly classify a set of measurement data as normal or malicious even when parts of the dataset are missing. We leverage such characteristics of the ML method in our IDS to detect attacks, especially the control mode related attacks on the PV system. We simulate the PV control modes to run power flow using the power simulator MATPOWER~\cite{zimmerman1997matpower} and generate exhaustive datsets for our study. We test several ML algorithms and compare their performance. %
The contributions of our work can be summarized as below:

\begin{itemize}
    \item We give one of the first studies where simultaneous attacks on PVs operating in different control modes are considered, under a strong attacker model which assumes capability of manipulating individual PV bus measurements to remain hidden. %

    \item For the various configurations, we create several baseline datasets using a standard test distribution network with real-world PV generation and load demand data (where small errors of the energy meter readings are unavoidable). We make these datasets publicly available to facilitate further research in this direction. %
    \item We evaluate the performance of many ML algorithms by conducting extensive experiments including cases where parts of measurement data are missing. The results show that the ML-based techniques, specifically multi-layer perceptron and random forest algorithms are effective and efficient in detecting attacks on various PV control modes (accuracy of around 95\% even with missing data).
\end{itemize}

\section{Literature Review}

Cyberattack scenarios in the broader context of smart grid have been well studied in the literature. Study on DER integrated grids has also been popular among the energy and cybersecurity research communities. %
Qi \textit{et al.}~\cite{qi2016cybersecurity} suggested a holistic framework for defence against cyberattcks in a network with high DER penetrations. The resilience design aspects at cyber, physical device and utility levels had been broadly discussed. In the same vein, Johnson \textit{et al.}~\cite{johnson2019power} proposed engineering design control of the DER devices and enclaving (i.e., segmentation) of the network with several DERs to enhance the cyberattack resiliency. In~\cite{de2019recommended}, the authors summarized the current industry practices for DER cybersecurity and also suggested some strategies to improve the security postures. Specifically for IDS, signature-based and behavioural-based solutions were studied in~\cite{jones2020implementation} to detect few types of attacks on PV inverters. For voltage control manipulation in low voltage distribution grid, \cite{kosek2016contextual} gives a contextual anomaly detection method based on an artificial neural network. Chavez \textit{et al.}~\cite{chavez2019hybrid} showed the importance of physical system features, in addition to the network traffic features, to identify certain types of attacks in a distribution network. They collect and use a combination of cyber-security data and power system and control information to propose a hybrid IDS for DER systems. Unsupervised~\cite{shilay2017catching} ML algorithms have been tested on a proposed edge-based IDS for PV system security. In~\cite{li2020data}, a more conventional approach of supervised learning ML methods were used to detect attacks, considering synchronized data from PV systems.

In a different direction, Li \textit{et al.}~\cite{li2019detection} use raw electrical waveform data and a high-dimensional data-driven approach to detect and identify cyber-physical attacks in distribution power grids with PVs. %
Another approach that has been widely studied in the power system context is the application of physics-based techniques for attack detection \cite{%
mashima2018securing, mashima2017artificial}. However, these solutions rely quite heavily on the availability and accuracy of all measurement data. For cases with missing measurements or meter reading errors, the performance and accuracy of such solutions would be limited. %

Overall, most of these works discussed here focus on single operating mode and the attacker doesn't particularly focus on remaining hidden. In contrast, here we consider simultaneous attacks on PVs operating in different control modes where the attacker manipulates PV bus measurements to remain hidden.

\section{Threat Model and Attack Modes}
\label{PVthreat}

Before we discuss in detail the possible threats on PV unit operations, we briefly describe the various possible operating modes of a PV unit. 
Once the attacker gets control, it actively tries to stay hidden by  manipulating the measurement data of the bus associated with the DER by sending data as if the DER were not attacked. As such, an analysis of individual PV bus measurements wouldn't reveal any attack.

\subsection{PV Operating Modes}
\label{PVmodes}
We consider three PV operating modes, namely, limit active power mode (Max P), Constant power factor (PF) mode, and voltage-reactive power mode (volt-var). Due to variability in solar irradiance, the active power output from the PV unit changes and it is limited by the capacity of PV unit and associated inverters.

\begin{figure}
 \centering
    \includegraphics[width=0.71\linewidth]{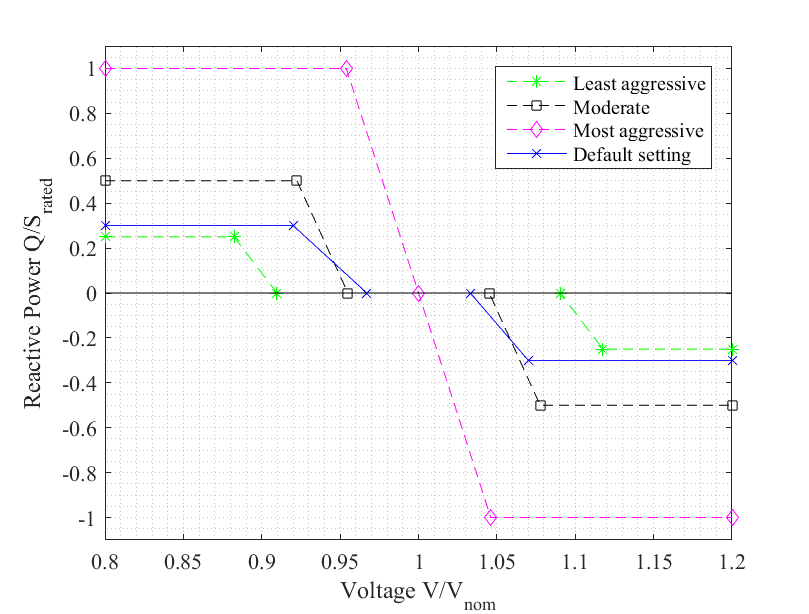}
  \caption{Volt-var standard characteristics\vspace{-3mm}}
  \label{fig:Voltvar}
\end{figure}

In \textit{limit active power mode (Max $P$)}, a DER is set to deliver a defined maximum amount of active power. 
In \textit{constant power factor (PF) mode}, the active power output is proportional to the reactive power output. Lastly, the  \textit{voltage-reactive power mode (volt-var mode)} of operation is an important regulation mode where the DER reactive power output is a function of the voltage at the point of common coupling (PCC) or the DER terminal for a standalone unit. PV inverter can be set to operate at any characteristics between the most and least aggressive curves defined as per UL 1741, and depicted in Fig.~\ref{fig:Voltvar}. Our default setting (blue %
line in Fig.~\ref{fig:Voltvar}) is as per the interconnection Rule 21 of California Public Utilities Commission, and also per the PV inverter application guide~\cite{PVINV}.

\subsection{PV Attack Modes}
\label{Attackmodes}
As the penetration of DERs including PVs into the grid is becoming high, any maloperation in the PV control modes would hamper the grid operation and may inflict economic loss. A direct attack on PV operation is to disconnect the unit from grid. An attacker may also manipulate the configuration (the control curve) of a PV inverter to change the amount of reactive power absorbed from or injected into the grid. This may incur additional system losses, affect the voltage stability and even permanently damage power system components. We briefly discuss about the possible attacks on PV systems. 

\noindent \textbf{Power curtailment attack:} When the DER operates in `limit active power mode', an adversary could lower the active power set point (or reduce it to zero) so that the actual power injected by the DER is lower than what it is capable of delivering. %
This type of attack may not have immediate significant impact on the grid, however, this type of attack is difficult to detect as the DER output is inherently variable and the attacker could hide within the fluctuating PV output.

\noindent \textbf{Disconnect attack:} Disconnect attack is also an attack on active power output (usually zero power output) of the DERs when the attacker is able to control a large number of PVs in the network. The disconnection of several DERs at once would increase the amount of power drawn from the grid, which may eventually lead to line overload, voltage and frequency fluctuations. The attack may also affect the grid planning as the DER owner would be unable to sell power to the utility. %

\noindent \textbf{Constant power factor (PF) mode attack:} An attacker might change the actual power factor setting to make it more capacitive or inductive resulting in poor voltage regulation, power quality and/or higher losses in the power system.

\noindent \textbf{Volt-var attack:} Smart DER inverters can be utilized to control voltage at certain point in the network (e.g., point of common coupling) by absorbing or injecting certain amount of reactive power. When an adversary gains knowledge of the control algorithm or the volt-var characteristic, (s)he can change the characteristic or even invert it to absorb or inject an arbitrary amount of reactive power. This would greatly hamper the voltage control application, and would affect the voltage level at the grid.

\noindent\textbf{Reverse power flow attack:} Modern households are equipped with smart appliances that adopt communication channel for remote control and monitoring. An attacker could exploit any weak links in the communication channel to turn the devices off. When the DER generation is high, such an event would result in supplying excess power to the grid instead of local consumption. If a large number of loads are under control of the attacker, the cumulative effect would impact the voltage regulation and protection coordination of the grid. 

\begin{figure}
 \centering
    \includegraphics[width=0.7\linewidth]{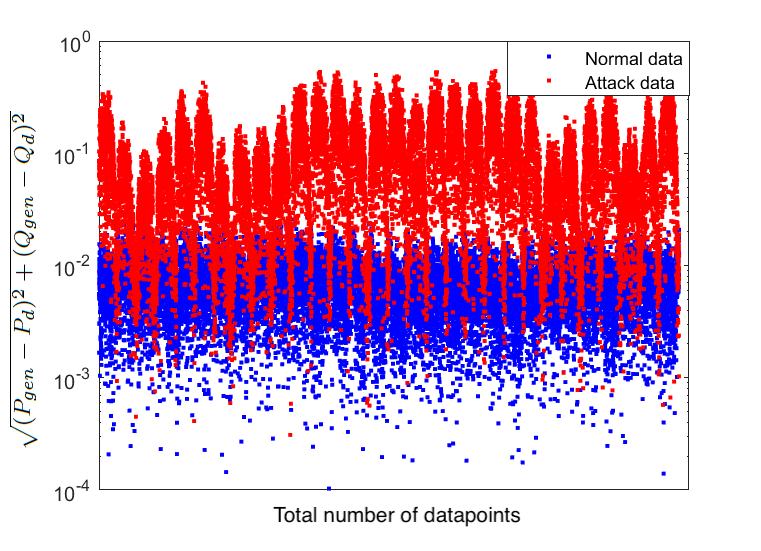}
  \caption{Plot for power flow test results}
  \label{fig:PFresult}
\end{figure}

\section{Physics-based vs ML-based Approach} \label{sec:discussion}
In this section, we discuss why we focus on ML-based detection of attacks on DER systems instead of physics-based method. From a given datapoint the objective of either approach is to determine the presence/absence of an attack. %
In real distribution networks, the differences in overall generation and consumption data denotes the inherent `loss' in the network, occuring primarily due to the impedance of the distribution lines. When the PV system is attacked and the associated PV system meter readings are manipulated (to maintain the pretence of non attack at individual level), the apparent overall loss value would be different from the actual system loss. Furthermore, small meter errors (maximum $\pm{1\%}$) are unavoidable and such cumulative errors of numerous meters would make it even harder to calculate the actual system losses, even if there is no attack. We evaluate the performance of physics-based study, i.e., power flow study, in detecting attacks on the PV system.

\begin{figure}
 \centering
    \includegraphics[width=0.75\linewidth]{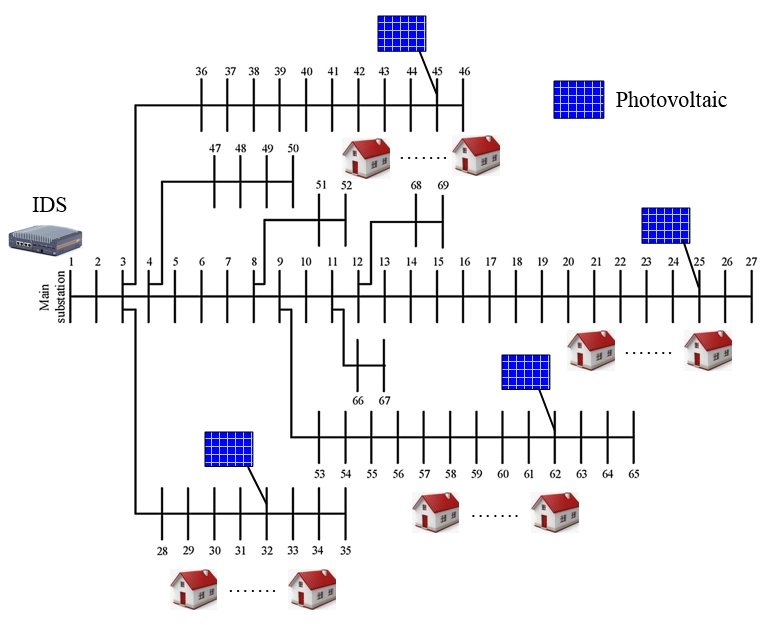}
  \caption{Adapted 69-bus distribution network}
  \label{fig:69bus}
\end{figure}

Fig~\ref{fig:PFresult} shows the plot of a form of network `losses' for the normal and attack datapoints. The attack cases consider PV control mode attacks, PV system meter data manipulation and inherent meter errors, while the normal datapoints have only the meter errors. As observed from the plot, a large number of normal and attack datapoints overlap, and it would not be easy to delimit an appropriate boundary between the two types of datapoints by some thresholding technique. Therefore, we resort to the supervised ML models that %
can learn from the available data and help develop a mechanism to identify attacks on the PV system.  %
The ML-based detector can also function with some missing measurement data in the system. The power flow study would give incorrect results or may even fail to converge if some measurement data are unavailable.

\section{ML-based IDS}
We make use of several well-known and popular machine learning algorithms/classifiers to evaluate the accuracy of those models as an IDS.  Most of the algorithms we consider here are simple supervised ML algorithms and once trained, for any given input can output whether or not there exists an attack.  %
As these algorithms are quite well-known and well-used in literature \cite{boser1992training, %
cheng2020powernet,shah2020comparative} we only provide a brief description. These classifiers were implemented in python using the sklearn library with a 5-fold cross validation. To make our results reproducible, we provide the tuned hyper-parameters for the generated datasets (in section \ref{sec:ml_datasets}) for each of the algorithms. 

\noindent\textbf{Logistic Regression (LR):}
LR is one of the simpler and widely used ML algorithms for binary classification and is typically used as a baseline for performance comparison. %
It estimates the relationship between the independent variables (features) and target binary variable (label). Based on a grid search, we obtain the optimal hyper-parameters for LR to be $C$=100, max\_iter=1000, penalty=`l1', and solver=`liblinear'.

\noindent\textbf{Support Vector Machine (SVM):} For SVM, %
the classification is performed by determining a hyperplane that differentiates two classes to the largest extent (or by a maximum margin). Our grid search determines the optimal parameters to be $C$=1000, gamma=1, and kernel=`rbf'.

\noindent \textbf{$k$ Nearest Neighbors (KNN):} KNN %
works on the basic assumption that similar things/data-points exist in close proximity of one another and the classification is based on majority voting by the neighbors. The optimal parameters obtained through a grid search in this case was metric=`minkowski', n\_neighbors=2, and weights=`uniform'.

\noindent \textbf{Gradient boosting (GBT):}  GBT creates the prediction model in a sequence of steps, where in each step, it minimizes a predefined loss function by adding a tree, while keeping the existing trees unchanged. %
The grid search in this case yielded learning\_rate=0.01, max\_depth=4, max\_features=`sqrt', n\_estimators=1750, random\_state=10, and subsample=1 as the optimal parameters.

\noindent\textbf{Random Forest (RF):} RF is another ensemble ML method consisting of many uncorrelated decision trees. For classification, the output of the RF classifier is the class selected by most trees. In this case, the grid search yields max\_depth=100, max\_features=3, min\_samples\_leaf=3, min\_samples\_split=12 as the set of optimal parameters.

\noindent\textbf{Multi-layer Perceptron (MLP):} MLP is a feed-forward artificial neural network based classifier where each neural network layer is fully connected to the following one. Neurons have nonlinear activation functions, except for those of the input layer. %
The optimal model was obtained after a grid search that gave the parameters activation=`tanh', alpha=0.0001, hidden\_layer\_sizes=(50, 100, 50), learning\_rate=`adaptive', max\_iter=1000, and solver=`adam'.

\section{Case Studies and Evaluations}
\label{evaluations}
We evaluate the performance of our proposed IDS on a standard distribution network where PV systems, domestic and small scale commercial consumers are connected. Real world data for PV generation and domestic household demands of every single minute over a period of one month has been considered for training and testing purpose of the several ML algorithms, and to assess the overall performance of the IDS. 

\subsection{Test Network}
We consider the 69-bus 12.66kV distribution network proposed by Baran and Wu~\cite{baran1989optimal}, and reproduced here in Fig.~\ref{fig:69bus}. The network has been widely adopted by the research community to study optimal placement of DERs and shunt capacitors~\cite{biswas2017multiobjective,almabsout2020hybrid}, as well as for optimal network reconfiguration~\cite{biswas2018distribution}. We use the default base configuration and branch data as in the original proposed network. Bus load data is altered as described herein.

\noindent\textbf{Load data:} For bus loading, we consider the variable domestic household consumption data provided in Smart$^*$%
~\cite{barker2012smart} dataset. Smart$^*$ contains every minute energy consumption data for three real homes - Home A, Home B and Home C, for several months and years. We use the data for the month of June in the year 2012. While assigning these data to the network buses, we consider that a minimum of 4 and maximum 10 houses (of any combination of home types selected randomly) are connected to each bus. While doing so, we also ensure that the network load at any moment does not exceed the total load originally proposed in~\cite{baran1989optimal}, i.e., 3802.19 kW and 2964.6 kvar. In our power system study, we ignore the branch-line or cable required to interconnect all houses connected to each bus, and assume that the load at the point of common coupling is balanced. Also, one smart meter shall be able to transmit energy consumption data of all the homes connected to a single bus. We do not consider any load at substation bus 1, bus 2, bus 3 (as in the original network), and also at the buses where PV plants are connected. A summary of the network is provided in Table~\ref{tbl:Testnet}. The average power factor for the household loads is considered to be 0.78 lagging, in line with the detailed study performed in~\cite{ahmad2019tariff}. We publicly share the synthesized time-series consumption data in GitHub repository\footnote{\url{https://github.com/suman-nus/Synthesized-DER-Data}}. 

\begin{table}[!htbp]
\centering
\caption{Summary of the Test Network}
\label{tbl:Testnet}
\begin{tabular}{|p{11mm}|p{10mm}|p{44mm}|}\hline
\textbf{Items} & \textbf{Quantity} & \textbf{Details} \\ \hline
Buses & 69 & Base voltage: 12.66 kV~\cite{baran1989optimal} \\ \hline
Branches & 73 & Open branches: 69 to 73; \newline Branch data~\cite{baran1989optimal} \\ \hline
PV units & 4 & Bus 25 (420 kW), Bus 32 (180 kW), \newline Bus 45 (330 kW), Bus 62 (390 kW) \\ \hline
Loads & - & Consumption data:~\cite{barker2012smart}; \newline No load: Buses 1, 2, 3, 25, 32, 45, 62 \\ \hline
\end{tabular}
\end{table}

\noindent\textbf{PV system data:} We consider 4 nos. PV plants in our test network. Singapore PV generation data for selected 4 plants are incorporated in our study. The data contains PV generation of every minute for the month of May in the year 2020 (data from 1st-30th May, 2020 are considered in study). In a distribution network, the location of the PV is decided based on the load demand at various buses of the network. In our test case, placement of the PV units closer to the farthest points of the radial feeders would provide voltage support, as well as would help reduce the network losses. Therefore, we consider the 4 PV units that are connected to bus 25, bus 32, bus 45 and bus 62 (refer to Fig.~\ref{fig:69bus}) with maximum kW capacities as mentioned in Table~\ref{tbl:Testnet}. Although the choice of PV locations in our experimental set up is somewhat arbitrary, finding optimal siting for the same is outside the scope of this work.

\subsection{Datasets for the Machine Learning Algorithm} \label{sec:ml_datasets}

We consider consumption and generation data per minute for 12 hours each day for a month-long period. The purpose of our study is to detect attacks on PV system control functions. Grid connected PVs could be turned off during night, and to have meaningful gain, the attack would likely be carried out during daytime when the PV generation is at substantial level. Now, out of approximately 21600 total datapoints over a month, 80\% of the datapoints are used to train the ML algorithm, while the remaining are utilized to test it. Both training and testing datasets have normal and attack datapoints. Normal datapoints are for the naturally variable demand data and the PV generation data. 3-different settings of the network, based on operating modes of the 4 PV units, are considered in our study as detailed in Table~\ref{tbl:Attackscene}.  %
For all the settings, the attack datapoints are created in the following manner.

\begin{itemize}
    \item 20\% of the total datapoints are randomly selected so that they can be corrupted to generate attack datapoints. 
    \item Each DER is assumed to operate in one of the 3 operating modes: constant power factor, limit active power (Max $P$) and volt-var control.
    \item One or multiple DERs, operating all in same or different modes, may be attacked. The attack scenarios considered in our study for different settings of the network are provided in Table~\ref{tbl:Attackscene}.
    \item In each attack datapoint either 1 PV, 2 PVs or all the 4 PVs are considered under attack (randomly chosen).
\end{itemize}

In the context of the PV attack modes discussed in section~\ref{Attackmodes}, it may be noted that the effect of power curtailment attack, disconnect attack and reverse power flow attack can be simulated with the change in setpoint of active power $P$ when the PV is operating in limit active power (i.e., Max $P$) mode. For both normal and attack datasets, each datapoint contains active power ($P$) and reactive power ($Q$) for all buses. We consider measurement error of up to $\pm{1\%}$. The standard IEEE C57.13.6 defines high accuracy classes of instrument transformers such as $0.15$S, $0.3$S, $0.5$S, etc. to have maximum errors of $\pm{0.15\%}$, $\pm{0.3\%}$ and $\pm{0.5\%}$, respectively. Moreover, metering instrument accuracy classes of 0.2, 0.5, etc. are also stipulated in ANSI standard C12.20. Therefore, our assumption of maximum error level of up to $\pm{1\%}$ is well justified if the metering instruments are of high accuracy class. In case of an attack on DER control mode, we presume that the attacker manipulates the DER settings (or control characteristics), as well as the measurement data of the bus associated with the DER as if the DER were not attacked. Therefore, the attacker can remain hidden in the system as the PV bus measurements won't reveal any anomaly. We add all $P$ and $Q$ received from all the load buses. In the dataset, we have a total of 6 features for the machine learning models: sum of all $P$ and $Q$ readings (say, $P_d$ and $Q_d$, respectively) obtained from load buses, substation bus-1 $P$ and $Q$ measurement data (say, $P_{gen}$ and $Q_{gen}$, respectively), and the differences of these values, i.e., $(P_{gen}-P_d)$ and $(Q_{gen}-Q_d)$.

\noindent \textbf{Missing Measurement data:} In a large distribution network numerous smart meters would send measurement data to an aggregator or a nearby substation. In the process, it is likely that some measurement data would be missed due to a faulty meter or a communication loss. We create datasets to account for such cases and evaluate the performance of our IDS on these datasets. In the measurement datasets, we consider 20\% of the total datapoints have missing data where in each datapoint, maximum 10\% of the bus data (both $P$ and $Q$ data for the buses) are missing. For example, if a measurement datapoint has meter readings from 60 buses of the network, readings of 6 buses would be missing (total 12 missing meter readings considering both $P$ and $Q$). Now, if we have a total of 1000 measurement datapoints, 200 of them would have such missing data. Similar to the earlier case, we sum all the available bus $P$ and $Q$ data with missing data treated as 0.

\begin{table*}[!htbp]
\centering
\caption{{PV Operating Modes and Attack Scenarios for Case Studies}}
\label{tbl:Attackscene}
\begin{tabular}{|c|cccc|c|}
\hline
\multirow{3}{*}{\begin{tabular}[c]{@{}c@{}}Network \\ Settings\end{tabular}} & \multicolumn{4}{c|}{Normal operations}                                                                     & \multirow{3}{*}{Attack Details}                                              \\ \cline{2-5}
                                                                             & \multicolumn{4}{c|}{PV Operating Modes}                                                                    &                                                                              \\ \cline{2-5}
                                                                             & \multicolumn{1}{c|}{PV1}      & \multicolumn{1}{c|}{PV2}           & \multicolumn{1}{c|}{PV3}   & PV4      &                                                                              \\ \hline
Setting 1                                                                    & \multicolumn{4}{c|}{All in constant PF = 1 mode}                                                           & PF changed to 0.8 (lag) or 0.8 (lead)                                        \\ \hline
Setting 2                                                                    & \multicolumn{4}{c|}{All in limit active power, i.e., Max P mode}                                           & MaxP is changed to any value between 0 to 0.8 × setpoint                     \\ \hline
Setting 3                                                                    & \multicolumn{4}{c|}{All in volt-var control (Default CA Rule 21) mode}                                     & Volt-var characteristic setpoints are changed to arbitrary values, inverted  \\ \hline
Setting 4                                                                    & \multicolumn{1}{c|}{Volt-var} & \multicolumn{1}{c|}{Constant PF = 1} & \multicolumn{1}{c|}{Max P} & Volt-var & Max P, PF setpoints and Volt-var characteristic changed as mentioned earlier \\ \hline
\end{tabular}
\end{table*}

\begin{table*}[!htbp]
\caption{Training/Testing Schemes}
\label{tbl:scheme}
\centering

\resizebox{2\columnwidth}{!}{\begin{tabular}{|c|c|c|}
\hline
 & Training Data Characteristics & Testing Data Characteristics \\ \hline
Scheme 1 & All measurements are available & All measurements are available \\ \hline
Scheme 2 & All measurements are available & With random missing measurements (max. 10\% of missing bus data) \\ \hline
Scheme 3 & With random missing measurements (max. 10\% of missing bus data) & With random missing measurements (max. 10\% of missing bus data) \\ \hline
\end{tabular}}
\end{table*}

\subsection{Performance of various ML algorithms}

To more accurately simulate different real world settings, we consider three training/testing schemes as shown in Table \ref{tbl:scheme} where we consider inputs with/without missing measurement data. We evaluate the performance of three schemes under different PV operating modes and attack scenarios using several different evaluation metrics. Thereafter, we discuss how the ML algorithms performs in these cases.

To get a holistic view of the performance, we calculate several different well known evaluation metrics like Accuracy, Precision, Recall, AUC (area under the ROC curve), PR AUC (precision recall curve) and Jaccard score. In the interest of space, we refer the reader to \cite{10.5555/1964882, jaccard} for details regarding the evaluation metrics. For each of the evaluation metrics, the possible values ranges from 0 to 1, with a value closer to 1 implying better performance. See Tables \ref{tab:fixed_pf_mode}, \ref{tab:max_p_mode}, \ref{tab:volt_var_mode}, and \ref{tab:all_modes} for the performance of the ML algorithms in each scenario.

We observe that the performance of all the algorithms are better when trained and tested on similar datasets (either with all available data, or with missing data) as compared to the case where training is done on all available data and testing is done with missing data. However, even in this case the ML algorithms, especially RF and MLP perform reasonably well.

Overall, the best performing algorithms in most cases are also the MLP algorithm and the RF algorithm. These dominate the top results (shown in bold font in Tables \ref{tab:fixed_pf_mode}, \ref{tab:max_p_mode}, \ref{tab:volt_var_mode}, and \ref{tab:all_modes}) or are a close second in most cases. The decision tree based algorithms (RF and GBT) perform in nearly identical fashion with RF holding the edge in many cases. However, the difference in many cases is quite insignificant and can possibly be attributed to fine grained hyper-parameter tuning.

\begin{table*}[]
\centering
\caption{Results of different ML algorithms for Network Setting 1 (all PVs are in Constant PF = 1 mode)}
\label{tab:fixed_pf_mode}
\resizebox{2\columnwidth}{!}{\begin{tabular}{|c|cccccc|cccccc|cccccc|}
\hline
\multirow{2}{*}{} & \multicolumn{6}{c|}{Scheme 1} & \multicolumn{6}{c|}{Scheme 2} & \multicolumn{6}{c|}{Scheme 3} \\ \hline
 & \multicolumn{1}{c|}{LR} & \multicolumn{1}{c|}{SVM} & \multicolumn{1}{c|}{KNN} & \multicolumn{1}{c|}{RF} & \multicolumn{1}{c|}{GBT} & MLP & \multicolumn{1}{c|}{LR} & \multicolumn{1}{c|}{SVM} & \multicolumn{1}{c|}{KNN} & \multicolumn{1}{c|}{RF} & \multicolumn{1}{c|}{GBT} & MLP & \multicolumn{1}{c|}{LR} & \multicolumn{1}{c|}{SVM} & \multicolumn{1}{c|}{KNN} & \multicolumn{1}{c|}{RF} & \multicolumn{1}{c|}{GBT} & MLP \\ \hline
Accuracy & \multicolumn{1}{c|}{0.859} & \multicolumn{1}{c|}{0.936} & \multicolumn{1}{c|}{0.929} & \multicolumn{1}{c|}{\textbf{0.963}} & \multicolumn{1}{c|}{\textbf{0.963}} & 0.962 & \multicolumn{1}{c|}{0.782} & \multicolumn{1}{c|}{0.788} & \multicolumn{1}{c|}{0.801} & \multicolumn{1}{c|}{\textbf{0.81}} & \multicolumn{1}{c|}{\textbf{0.81}} & 0.809 & \multicolumn{1}{c|}{0.85} & \multicolumn{1}{c|}{0.899} & \multicolumn{1}{c|}{0.902} & \multicolumn{1}{c|}{0.943} & \multicolumn{1}{c|}{0.943} & \textbf{0.949} \\ \hline
Precision & \multicolumn{1}{c|}{\textbf{1}} & \multicolumn{1}{c|}{\textbf{1}} & \multicolumn{1}{c|}{0.978} & \multicolumn{1}{c|}{0.994} & \multicolumn{1}{c|}{0.992} & 0.985 & \multicolumn{1}{c|}{0.448} & \multicolumn{1}{c|}{0.48} & \multicolumn{1}{c|}{0.502} & \multicolumn{1}{c|}{\textbf{0.516}} & \multicolumn{1}{c|}{0.515} & 0.513 & \multicolumn{1}{c|}{0.962} & \multicolumn{1}{c|}{0.935} & \multicolumn{1}{c|}{0.93} & \multicolumn{1}{c|}{0.954} & \multicolumn{1}{c|}{\textbf{0.957}} & 0.95 \\ \hline
Recall & \multicolumn{1}{c|}{0.296} & \multicolumn{1}{c|}{0.681} & \multicolumn{1}{c|}{0.662} & \multicolumn{1}{c|}{0.821} & \multicolumn{1}{c|}{0.822} & \textbf{0.823} & \multicolumn{1}{c|}{0.38} & \multicolumn{1}{c|}{0.726} & \multicolumn{1}{c|}{0.695} & \multicolumn{1}{c|}{0.854} & \multicolumn{1}{c|}{\textbf{0.855}} & 0.845 & \multicolumn{1}{c|}{0.26} & \multicolumn{1}{c|}{0.531} & \multicolumn{1}{c|}{0.554} & \multicolumn{1}{c|}{0.751} & \multicolumn{1}{c|}{0.75} & \textbf{0.786} \\ \hline
F1 Score & \multicolumn{1}{c|}{0.457} & \multicolumn{1}{c|}{0.81} & \multicolumn{1}{c|}{0.79} & \multicolumn{1}{c|}{\textbf{0.899}} & \multicolumn{1}{c|}{\textbf{0.899}} & 0.897 & \multicolumn{1}{c|}{0.411} & \multicolumn{1}{c|}{0.578} & \multicolumn{1}{c|}{0.583} & \multicolumn{1}{c|}{\textbf{0.643}} & \multicolumn{1}{c|}{\textbf{0.643}} & 0.638 & \multicolumn{1}{c|}{0.409} & \multicolumn{1}{c|}{0.677} & \multicolumn{1}{c|}{0.694} & \multicolumn{1}{c|}{0.841} & \multicolumn{1}{c|}{0.841} & \textbf{0.86} \\ \hline
AUC & \multicolumn{1}{c|}{0.713} & \multicolumn{1}{c|}{0.933} & \multicolumn{1}{c|}{0.871} & \multicolumn{1}{c|}{0.957} & \multicolumn{1}{c|}{0.963} & \textbf{0.965} & \multicolumn{1}{c|}{0.673} & \multicolumn{1}{c|}{0.827} & \multicolumn{1}{c|}{0.792} & \multicolumn{1}{c|}{\textbf{0.915}} & \multicolumn{1}{c|}{0.812} & 0.836 & \multicolumn{1}{c|}{0.705} & \multicolumn{1}{c|}{0.889} & \multicolumn{1}{c|}{0.843} & \multicolumn{1}{c|}{0.94} & \multicolumn{1}{c|}{0.942} & \textbf{0.952} \\ \hline
PR AUC & \multicolumn{1}{c|}{0.697} & \multicolumn{1}{c|}{0.909} & \multicolumn{1}{c|}{0.852} & \multicolumn{1}{c|}{0.929} & \multicolumn{1}{c|}{0.932} & \textbf{0.935} & \multicolumn{1}{c|}{0.442} & \multicolumn{1}{c|}{0.636} & \multicolumn{1}{c|}{0.636} & \multicolumn{1}{c|}{\textbf{0.823}} & \multicolumn{1}{c|}{0.371} & 0.44 & \multicolumn{1}{c|}{0.63} & \multicolumn{1}{c|}{0.816} & \multicolumn{1}{c|}{0.792} & \multicolumn{1}{c|}{0.892} & \multicolumn{1}{c|}{0.894} & \textbf{0.908} \\ \hline
Jaccard & \multicolumn{1}{c|}{0.296} & \multicolumn{1}{c|}{0.681} & \multicolumn{1}{c|}{0.653} & \multicolumn{1}{c|}{\textbf{0.817}} & \multicolumn{1}{c|}{0.816} & 0.813 & \multicolumn{1}{c|}{0.259} & \multicolumn{1}{c|}{0.407} & \multicolumn{1}{c|}{0.411} & \multicolumn{1}{c|}{\textbf{0.474}} & \multicolumn{1}{c|}{0.473} & 0.469 & \multicolumn{1}{c|}{0.257} & \multicolumn{1}{c|}{0.512} & \multicolumn{1}{c|}{0.532} & \multicolumn{1}{c|}{0.725} & \multicolumn{1}{c|}{0.726} & \textbf{0.755} \\ \hline
\end{tabular}}
\end{table*}

\begin{table*}[]
\centering
\caption{Results of different ML algorithms for Network Setting 2 (all PVs are in Max $P$ mode)}
\label{tab:max_p_mode}
\resizebox{2\columnwidth}{!}{\begin{tabular}{|c|cccccc|cccccc|cccccc|}
\hline
 & \multicolumn{6}{c|}{Scheme 1} & \multicolumn{6}{c|}{Scheme 2} & \multicolumn{6}{c|}{Scheme 3} \\
 \hline
 & \multicolumn{1}{c|}{LR} & \multicolumn{1}{c|}{SVM} & \multicolumn{1}{c|}{KNN} & \multicolumn{1}{c|}{RF} & \multicolumn{1}{c|}{GBT} & MLP & \multicolumn{1}{c|}{LR} & \multicolumn{1}{c|}{SVM} & \multicolumn{1}{c|}{KNN} & \multicolumn{1}{c|}{RF} & \multicolumn{1}{c|}{GBT} & MLP & \multicolumn{1}{c|}{LR} & \multicolumn{1}{c|}{SVM} & \multicolumn{1}{c|}{KNN} & \multicolumn{1}{c|}{RF} & \multicolumn{1}{c|}{GBT} & MLP \\ \hline
Accuracy & \multicolumn{1}{c|}{0.956} & \multicolumn{1}{c|}{0.963} & \multicolumn{1}{c|}{0.948} & \multicolumn{1}{c|}{0.968} & \multicolumn{1}{c|}{0.968} & \textbf{0.969} & \multicolumn{1}{c|}{0.837} & \multicolumn{1}{c|}{0.842} & \multicolumn{1}{c|}{0.841} & \multicolumn{1}{c|}{0.845} & \multicolumn{1}{c|}{0.844} & \textbf{0.846} & \multicolumn{1}{c|}{0.941} & \multicolumn{1}{c|}{0.944} & \multicolumn{1}{c|}{0.931} & \multicolumn{1}{c|}{0.958} & \multicolumn{1}{c|}{0.957} & \textbf{0.958} \\ \hline
Precision & \multicolumn{1}{c|}{0.949} & \multicolumn{1}{c|}{\textbf{1}} & \multicolumn{1}{c|}{0.989} & \multicolumn{1}{c|}{0.991} & \multicolumn{1}{c|}{0.991} & \textbf{1} & \multicolumn{1}{c|}{0.56} & \multicolumn{1}{c|}{0.57} & \multicolumn{1}{c|}{0.573} & \multicolumn{1}{c|}{0.573} & \multicolumn{1}{c|}{0.572} & \textbf{0.574} & \multicolumn{1}{c|}{0.966} & \multicolumn{1}{c|}{0.991} & \multicolumn{1}{c|}{0.964} & \multicolumn{1}{c|}{0.982} & \multicolumn{1}{c|}{0.982} & \textbf{0.999} \\ \hline
Recall & \multicolumn{1}{c|}{0.824} & \multicolumn{1}{c|}{0.816} & \multicolumn{1}{c|}{0.75} & \multicolumn{1}{c|}{0.849} & \multicolumn{1}{c|}{\textbf{0.85}} & 0.844 & \multicolumn{1}{c|}{0.858} & \multicolumn{1}{c|}{0.85} & \multicolumn{1}{c|}{0.797} & \multicolumn{1}{c|}{0.877} & \multicolumn{1}{c|}{0.879} & \textbf{0.88} & \multicolumn{1}{c|}{0.728} & \multicolumn{1}{c|}{0.725} & \multicolumn{1}{c|}{0.68} & \multicolumn{1}{c|}{\textbf{0.806}} & \multicolumn{1}{c|}{0.8} & 0.791 \\ \hline
F1 Score & \multicolumn{1}{c|}{0.882} & \multicolumn{1}{c|}{0.899} & \multicolumn{1}{c|}{0.853} & \multicolumn{1}{c|}{0.914} & \multicolumn{1}{c|}{\textbf{0.915}} & \textbf{0.915} & \multicolumn{1}{c|}{0.677} & \multicolumn{1}{c|}{0.682} & \multicolumn{1}{c|}{0.667} & \multicolumn{1}{c|}{0.693} & \multicolumn{1}{c|}{0.693} & \textbf{0.695} & \multicolumn{1}{c|}{0.831} & \multicolumn{1}{c|}{0.837} & \multicolumn{1}{c|}{0.797} & \multicolumn{1}{c|}{\textbf{0.885}} & \multicolumn{1}{c|}{0.882} & 0.883 \\ \hline
AUC & \multicolumn{1}{c|}{0.975} & \multicolumn{1}{c|}{0.984} & \multicolumn{1}{c|}{0.91} & \multicolumn{1}{c|}{0.983} & \multicolumn{1}{c|}{0.984} & \textbf{0.986} & \multicolumn{1}{c|}{0.892} & \multicolumn{1}{c|}{0.89} & \multicolumn{1}{c|}{0.849} & \multicolumn{1}{c|}{\textbf{0.943}} & \multicolumn{1}{c|}{0.854} & 0.892 & \multicolumn{1}{c|}{0.944} & \multicolumn{1}{c|}{0.952} & \multicolumn{1}{c|}{0.891} & \multicolumn{1}{c|}{0.971} & \multicolumn{1}{c|}{0.971} & \textbf{0.978} \\ \hline
PR AUC & \multicolumn{1}{c|}{0.945} & \multicolumn{1}{c|}{0.958} & \multicolumn{1}{c|}{0.902} & \multicolumn{1}{c|}{0.959} & \multicolumn{1}{c|}{0.958} & \textbf{0.962} & \multicolumn{1}{c|}{0.648} & \multicolumn{1}{c|}{0.703} & \multicolumn{1}{c|}{0.713} & \multicolumn{1}{c|}{\textbf{0.855}} & \multicolumn{1}{c|}{0.431} & 0.531 & \multicolumn{1}{c|}{0.906} & \multicolumn{1}{c|}{0.917} & \multicolumn{1}{c|}{0.863} & \multicolumn{1}{c|}{0.936} & \multicolumn{1}{c|}{0.935} & \textbf{0.946} \\ \hline
Jaccard & \multicolumn{1}{c|}{0.79} & \multicolumn{1}{c|}{0.816} & \multicolumn{1}{c|}{0.744} & \multicolumn{1}{c|}{0.842} & \multicolumn{1}{c|}{0.843} & \textbf{0.844} & \multicolumn{1}{c|}{0.512} & \multicolumn{1}{c|}{0.518} & \multicolumn{1}{c|}{0.5} & \multicolumn{1}{c|}{0.53} & \multicolumn{1}{c|}{0.53} & \textbf{0.533} & \multicolumn{1}{c|}{0.71} & \multicolumn{1}{c|}{0.72} & \multicolumn{1}{c|}{0.663} & \multicolumn{1}{c|}{\textbf{0.794}} & \multicolumn{1}{c|}{0.788} & 0.79 \\ \hline
\end{tabular}}
\end{table*}

\begin{table*}[]
\centering
\caption{Results of different ML algorithms for Network Setting 3 (all PVs are in volt-var control mode)}
\label{tab:volt_var_mode}
\resizebox{2\columnwidth}{!}{\begin{tabular}{|c|cccccc|cccccc|cccccc|}
\hline
\multirow{2}{*}{} & \multicolumn{6}{c|}{Scheme 1} & \multicolumn{6}{c|}{Scheme 2} & \multicolumn{6}{c|}{Scheme 3} \\ \hline
 & \multicolumn{1}{c|}{LR} & \multicolumn{1}{c|}{SVM} & \multicolumn{1}{c|}{KNN} & \multicolumn{1}{c|}{RF} & \multicolumn{1}{c|}{GBT} & MLP & \multicolumn{1}{c|}{LR} & \multicolumn{1}{c|}{SVM} & \multicolumn{1}{c|}{KNN} & \multicolumn{1}{c|}{RF} & \multicolumn{1}{c|}{GBT} & MLP & \multicolumn{1}{c|}{LR} & \multicolumn{1}{c|}{SVM} & \multicolumn{1}{c|}{KNN} & \multicolumn{1}{c|}{RF} & \multicolumn{1}{c|}{GBT} & MLP \\ \hline
Accuracy & \multicolumn{1}{c|}{0.836} & \multicolumn{1}{c|}{0.932} & \multicolumn{1}{c|}{0.915} & \multicolumn{1}{c|}{0.96} & \multicolumn{1}{c|}{0.958} & \textbf{0.962} & \multicolumn{1}{c|}{0.832} & \multicolumn{1}{c|}{0.782} & \multicolumn{1}{c|}{0.794} & \multicolumn{1}{c|}{\textbf{0.801}} & \multicolumn{1}{c|}{0.8} & \textbf{0.801} & \multicolumn{1}{c|}{0.828} & \multicolumn{1}{c|}{0.889} & \multicolumn{1}{c|}{0.887} & \multicolumn{1}{c|}{0.937} & \multicolumn{1}{c|}{0.934} & \textbf{0.942} \\ \hline
Precision & \multicolumn{1}{c|}{\textbf{1}} & \multicolumn{1}{c|}{0.993} & \multicolumn{1}{c|}{0.949} & \multicolumn{1}{c|}{0.987} & \multicolumn{1}{c|}{0.978} & 0.996 & \multicolumn{1}{c|}{\textbf{1}} & \multicolumn{1}{c|}{0.47} & \multicolumn{1}{c|}{0.488} & \multicolumn{1}{c|}{0.502} & \multicolumn{1}{c|}{0.5} & 0.502 & \multicolumn{1}{c|}{\textbf{1}} & \multicolumn{1}{c|}{0.951} & \multicolumn{1}{c|}{0.884} & \multicolumn{1}{c|}{0.951} & \multicolumn{1}{c|}{0.948} & 0.961 \\ \hline
Recall & \multicolumn{1}{c|}{0.18} & \multicolumn{1}{c|}{0.665} & \multicolumn{1}{c|}{0.607} & \multicolumn{1}{c|}{0.81} & \multicolumn{1}{c|}{0.809} & \textbf{0.815} & \multicolumn{1}{c|}{0.162} & \multicolumn{1}{c|}{0.698} & \multicolumn{1}{c|}{0.609} & \multicolumn{1}{c|}{0.816} & \multicolumn{1}{c|}{\textbf{0.82}} & 0.812 & \multicolumn{1}{c|}{0.142} & \multicolumn{1}{c|}{0.469} & \multicolumn{1}{c|}{0.502} & \multicolumn{1}{c|}{0.72} & \multicolumn{1}{c|}{0.71} & \textbf{0.741} \\ \hline
F1 Score & \multicolumn{1}{c|}{0.306} & \multicolumn{1}{c|}{0.796} & \multicolumn{1}{c|}{0.74} & \multicolumn{1}{c|}{0.89} & \multicolumn{1}{c|}{0.886} & \textbf{0.896} & \multicolumn{1}{c|}{0.279} & \multicolumn{1}{c|}{0.562} & \multicolumn{1}{c|}{0.542} & \multicolumn{1}{c|}{\textbf{0.621}} & \multicolumn{1}{c|}{\textbf{0.621}} & 0.62 & \multicolumn{1}{c|}{0.249} & \multicolumn{1}{c|}{0.628} & \multicolumn{1}{c|}{0.64} & \multicolumn{1}{c|}{0.82} & \multicolumn{1}{c|}{0.812} & \textbf{0.837} \\ \hline
AUC & \multicolumn{1}{c|}{0.69} & \multicolumn{1}{c|}{0.935} & \multicolumn{1}{c|}{0.862} & \multicolumn{1}{c|}{0.958} & \multicolumn{1}{c|}{0.962} & \textbf{0.964} & \multicolumn{1}{c|}{0.68} & \multicolumn{1}{c|}{0.822} & \multicolumn{1}{c|}{0.768} & \multicolumn{1}{c|}{\textbf{0.91}} & \multicolumn{1}{c|}{0.827} & 0.887 & \multicolumn{1}{c|}{0.681} & \multicolumn{1}{c|}{0.802} & \multicolumn{1}{c|}{0.809} & \multicolumn{1}{c|}{0.938} & \multicolumn{1}{c|}{0.938} & \textbf{0.949} \\ \hline
PR AUC & \multicolumn{1}{c|}{0.662} & \multicolumn{1}{c|}{0.904} & \multicolumn{1}{c|}{0.824} & \multicolumn{1}{c|}{0.93} & \multicolumn{1}{c|}{0.935} & \textbf{0.936} & \multicolumn{1}{c|}{0.643} & \multicolumn{1}{c|}{0.626} & \multicolumn{1}{c|}{0.597} & \multicolumn{1}{c|}{\textbf{0.813}} & \multicolumn{1}{c|}{0.396} & 0.745 & \multicolumn{1}{c|}{0.643} & \multicolumn{1}{c|}{0.742} & \multicolumn{1}{c|}{0.733} & \multicolumn{1}{c|}{0.886} & \multicolumn{1}{c|}{0.881} & \textbf{0.904} \\ \hline
Jaccard & \multicolumn{1}{c|}{0.18} & \multicolumn{1}{c|}{0.662} & \multicolumn{1}{c|}{0.588} & \multicolumn{1}{c|}{0.802} & \multicolumn{1}{c|}{0.795} & \textbf{0.812} & \multicolumn{1}{c|}{0.162} & \multicolumn{1}{c|}{0.39} & \multicolumn{1}{c|}{0.372} & \multicolumn{1}{c|}{\textbf{0.451}} & \multicolumn{1}{c|}{\textbf{0.451}} & 0.45 & \multicolumn{1}{c|}{0.142} & \multicolumn{1}{c|}{0.458} & \multicolumn{1}{c|}{0.471} & \multicolumn{1}{c|}{0.695} & \multicolumn{1}{c|}{0.683} & \textbf{0.719} \\ \hline
\end{tabular}}
\end{table*}

\begin{table*}[]
\centering
\caption{Results of different ML algorithms for Network Setting 4 (PVs are in different operating modes).}
\label{tab:all_modes}
\resizebox{2\columnwidth}{!}{\begin{tabular}{|c|cccccc|cccccc|cccccc|}
\hline
\multirow{2}{*}{} & \multicolumn{6}{c|}{Scheme 1} & \multicolumn{6}{c|}{Scheme 2} & \multicolumn{6}{c|}{Scheme 3} \\\hline
 & \multicolumn{1}{c|}{LR} & \multicolumn{1}{c|}{SVM} & \multicolumn{1}{c|}{KNN} & \multicolumn{1}{c|}{RF} & \multicolumn{1}{c|}{GBT} & MLP & \multicolumn{1}{c|}{LR} & \multicolumn{1}{c|}{SVM} & \multicolumn{1}{c|}{KNN} & \multicolumn{1}{c|}{RF} & \multicolumn{1}{c|}{GBT} & MLP & \multicolumn{1}{c|}{LR} & \multicolumn{1}{c|}{SVM} & \multicolumn{1}{c|}{KNN} & \multicolumn{1}{c|}{RF} & \multicolumn{1}{c|}{GBT} & MLP \\ \hline
Accuracy & \multicolumn{1}{c|}{0.889} & \multicolumn{1}{c|}{0.942} & \multicolumn{1}{c|}{0.924} & \multicolumn{1}{c|}{\textbf{0.964}} & \multicolumn{1}{c|}{0.963} & 0.962 & \multicolumn{1}{c|}{0.777} & \multicolumn{1}{c|}{0.801} & \multicolumn{1}{c|}{0.802} & \multicolumn{1}{c|}{\textbf{0.814}} & \multicolumn{1}{c|}{0.813} & 0.813 & \multicolumn{1}{c|}{0.84} & \multicolumn{1}{c|}{0.889} & \multicolumn{1}{c|}{0.885} & \multicolumn{1}{c|}{0.923} & \multicolumn{1}{c|}{0.917} & \textbf{0.931} \\ \hline
Precision & \multicolumn{1}{c|}{0.946} & \multicolumn{1}{c|}{\textbf{0.997}} & \multicolumn{1}{c|}{0.975} & \multicolumn{1}{c|}{0.985} & \multicolumn{1}{c|}{0.982} & 0.976 & \multicolumn{1}{c|}{0.453} & \multicolumn{1}{c|}{0.502} & \multicolumn{1}{c|}{0.504} & \multicolumn{1}{c|}{\textbf{0.52}} & \multicolumn{1}{c|}{0.519} & \textbf{0.52} & \multicolumn{1}{c|}{0.917} & \multicolumn{1}{c|}{\textbf{0.949}} & \multicolumn{1}{c|}{0.912} & \multicolumn{1}{c|}{0.903} & \multicolumn{1}{c|}{0.909} & 0.908 \\ \hline
Recall & \multicolumn{1}{c|}{0.469} & \multicolumn{1}{c|}{0.712} & \multicolumn{1}{c|}{0.635} & \multicolumn{1}{c|}{\textbf{0.831}} & \multicolumn{1}{c|}{0.829} & 0.83 & \multicolumn{1}{c|}{0.562} & \multicolumn{1}{c|}{0.785} & \multicolumn{1}{c|}{0.697} & \multicolumn{1}{c|}{\textbf{0.871}} & \multicolumn{1}{c|}{0.869} & 0.861 & \multicolumn{1}{c|}{0.218} & \multicolumn{1}{c|}{0.471} & \multicolumn{1}{c|}{0.469} & \multicolumn{1}{c|}{0.691} & \multicolumn{1}{c|}{0.647} & \textbf{0.729} \\ \hline
F1 Score & \multicolumn{1}{c|}{0.628} & \multicolumn{1}{c|}{0.831} & \multicolumn{1}{c|}{0.769} & \multicolumn{1}{c|}{\textbf{0.902}} & \multicolumn{1}{c|}{0.899} & 0.897 & \multicolumn{1}{c|}{0.502} & \multicolumn{1}{c|}{0.613} & \multicolumn{1}{c|}{0.585} & \multicolumn{1}{c|}{\textbf{0.651}} & \multicolumn{1}{c|}{0.65} & 0.648 & \multicolumn{1}{c|}{0.353} & \multicolumn{1}{c|}{0.629} & \multicolumn{1}{c|}{0.62} & \multicolumn{1}{c|}{0.783} & \multicolumn{1}{c|}{0.756} & \textbf{0.809} \\ \hline
AUC & \multicolumn{1}{c|}{0.809} & \multicolumn{1}{c|}{0.952} & \multicolumn{1}{c|}{0.869} & \multicolumn{1}{c|}{0.97} & \multicolumn{1}{c|}{\textbf{0.973}} & \textbf{0.973} & \multicolumn{1}{c|}{0.731} & \multicolumn{1}{c|}{0.846} & \multicolumn{1}{c|}{0.796} & \multicolumn{1}{c|}{\textbf{0.862}} & \multicolumn{1}{c|}{0.832} & 0.853 & \multicolumn{1}{c|}{0.719} & \multicolumn{1}{c|}{0.812} & \multicolumn{1}{c|}{0.805} & \multicolumn{1}{c|}{0.945} & \multicolumn{1}{c|}{0.937} & \textbf{0.951} \\ \hline
PR AUC & \multicolumn{1}{c|}{0.725} & \multicolumn{1}{c|}{0.92} & \multicolumn{1}{c|}{0.847} & \multicolumn{1}{c|}{0.943} & \multicolumn{1}{c|}{0.943} & \textbf{0.946} & \multicolumn{1}{c|}{0.388} & \multicolumn{1}{c|}{0.656} & \multicolumn{1}{c|}{0.64} & \multicolumn{1}{c|}{\textbf{0.645}} & \multicolumn{1}{c|}{0.399} & 0.539 & \multicolumn{1}{c|}{0.596} & \multicolumn{1}{c|}{0.746} & \multicolumn{1}{c|}{0.734} & \multicolumn{1}{c|}{0.882} & \multicolumn{1}{c|}{0.862} & \textbf{0.896} \\ \hline
Jaccard & \multicolumn{1}{c|}{0.457} & \multicolumn{1}{c|}{0.71} & \multicolumn{1}{c|}{0.625} & \multicolumn{1}{c|}{\textbf{0.821}} & \multicolumn{1}{c|}{0.817} & 0.813 & \multicolumn{1}{c|}{0.335} & \multicolumn{1}{c|}{0.441} & \multicolumn{1}{c|}{0.413} & \multicolumn{1}{c|}{\textbf{0.483}} & \multicolumn{1}{c|}{0.482} & 0.48 & \multicolumn{1}{c|}{0.214} & \multicolumn{1}{c|}{0.459} & \multicolumn{1}{c|}{0.449} & \multicolumn{1}{c|}{0.644} & \multicolumn{1}{c|}{0.608} & \textbf{0.679} \\ \hline
\end{tabular}}
\end{table*}

\section{Conclusion}
In this work, we present an ML-based IDS solution for various possible attack modes on PV system controls. For study purpose, we develop a test distribution network incorporating real-world PV generation data and consumption data. %
We observe that the MLP and the RF algorithms exhibit the most superior performance with attack detection accuracy up to around 95\% for most scenarios. We believe that this obtained accuracy is sufficient for most cases and more sophisticated ML algorithms like RNN and transformers can perform even better at the cost of additional time and resources for training. In general, the ML-based approach can detect even a small attack that does not significantly impact the grid operation. %

\section*{Acknowledgements}

This research is supported in part by the National Research Foundation, Singapore, under its National Satellite of Excellence Programme “Design Science and Technology for Secure Critical Infrastructure” (Award Number: NSoE\_DeST-SCI2019-0008), and in part by the National Research Foundation, Prime Minister’s Office, Singapore under its Campus for Research Excellence and Technological Enterprise (CREATE) programme, and in part by the SUTD Start-up Research Grant (SRG Award No: SRG ISTD 2020 157). Any opinions, findings and conclusions or recommendations expressed in this material are those of the author(s) and do not reflect the views of National Research Foundation, Singapore.

\bibliographystyle{IEEEtran}
\begin{small}
\bibliography{ref}
\end{small}
\end{document}